\documentclass{article}
\title{Improved extended Hamiltonian and search for local symmetries}
\author{A. A. Deriglazov\footnote{alexei.deriglazov@ufjf.edu.br ~ On leave of
absence from Dept. Math. Phys., Tomsk Polytechnical University,
Tomsk, Russia.}}
\date{Dept. de Matematica, ICE, Universidade Federal de Juiz de Fora,\\
MG, Brazil; \\
and \\
LAFEX - CBPF/MCT, Rio de Janeiro, RJ, Brazil.}

\begin{document}
\maketitle
\large

\begin{abstract}
We analyze a structure of the singular Lagrangian $L$ with first and second class constraints of an arbitrary stage. We show that there exist an equivalent Lagrangian (called the extended Lagrangian $\tilde L$) that generates all the original constraints on second stage of the Dirac-Bergmann procedure. The extended Lagrangian is obtained in closed form through the initial one. The formalism implies an extension of the original configuration space by auxiliary variables. Some of them are identified with gauge fields supplying local symmetries of $\tilde L$.
As an application of the formalism, we found closed expression for the gauge generators of $\tilde L$ through the first class constraints. It turns out to be much more easy task as those for $L$. All the first class constraints of $L$ turn out to be the gauge symmetry generators of $\tilde L$. By this way, local symmetries of $L$ with higher order derivatives of the local parameters decompose into a sum of the gauge symmetries of $\tilde L$. It proves the Dirac conjecture in the Lagrangian framework.
\end{abstract}

\noindent
%{\bf PAC codes:} \\
%{\bf Keywords:} Constrained systems, Local symmetries, Gauge theories

\section{Introduction}
Dirac-Bergmann algorithm proves to be a principal tool for analysis of various field and particle theories with local (gauge) symmetries, and, more generally, of any theory constructed on the base of singular Lagrangian. While it has a solid mathematical ground and a well established interpretation [1-4], some problems within the formalism remain under investigation [5-18]. The aim of this work is to reveal one of the long standing problems, concerning the proper interpretation and treatment of so called extended Hamiltonian formulation of the singular system.

In the Hamiltonian framework, possible motions of the singular system are restricted to lie on some surface of a phase space. Algebraic
equations of the surface (Dirac constraints) can be revealed in the course of the Dirac-Bergmann  procedure, the latter in general case requires a number of stages. According to the order of appearance, the constraints are called primary, second-stage, ... , N-th stage constraints. All the constraints, beside the primary ones are called the higher-stage constraints and are denoted collectively $T_a$.
The basic object of the Hamiltonian formulation turns out to be the complete Hamiltonian $H$ $=$ $H_0$ $+$ $v^\alpha\Phi_\alpha$. Here $H_0$ is the Hamiltonian, $v^\alpha$ represents primarily inexpressible velocities [3], and $\Phi_\alpha$ are primary constraints.
The extended Hamiltonian is constructed adding by hand the higher stage constraints with the multipliers $\lambda^a$: $H_{ext}$ $\equiv$ $H$ $+$ $\lambda^aT_a$.
The Hamiltonian equations following from $H_{ext}$ involve the extra terms with derivatives of $T_a$ and hence are different from the equations obtained from $H$. Nevertheless, a detailed analysis in special basis on the phase space shows that physical sectors of the two formulations are equivalent [3].

All the constraints enter into $H_{ext}$ in the manifest form. By this reason, the extended Hamiltonian turns out to be a very useful tool for the analysis of both the general structure [3] and local symmetries [4, 5] of the singular theory. At the same time, since the higher stage constraints have been added by hand, the origin of the extended Hamiltonian and its proper interpretation in the Dirac-Bergmann framework remain somewhat mysterious. In particular, $H_{ext}$ cannot be treated as the complete Hamiltonian generated by some Lagrangian (see Sect. 2 for details). So one asks whether it is possible to construct an equivalent Lagrangian formulation that would generate the complete Hamiltonian of the same structure as $H_{ext}$. We solve this problem in the Section 3.

For the case of first class constraints, the problem has been discussed in the recent work [12]. Here we generalize this analysis to an arbitrary case, with first and second class constraints up to $N$-th stage presented in the original formulation $L$. We present an improvement of the extended Hamiltonian formalism according to the following scheme. Starting from the initial Lagrangian $L$ (provided all its constraints are known), we work out an equivalent Lagrangian $\tilde L$ called the extended Lagrangian. It is obtained in a closed form in terms of the quantities of  initial formulation (see Eq. (\ref{13}) below). Due to the equivalence of $L$ and $\tilde L$, it is matter of convenience what formulation is used to describe the theory under consideration.

By construction, all the Lagrangian counterparts of the higher-stage constraints $T_a$ enter into $\tilde L$ in the manifest form, see the last term in Eq. (\ref{13}). The complete Hamiltonian $\tilde H$ generated by $\tilde L$ has the same structure as $H_{ext}$. So, the improved formalism maintains all the advantages of the extended Hamiltonian formalism. Besides, since it originates from the Lagrangian, all the quantities appearing in the formalism have clear meaning in the Dirac framework.

We explore the extended Lagrangian formulation to resolve another long standing problem concerning search for constructive procedure that would give local symmetries of a given Lagrangian action [4-16]. It is well known that in a singular theory there exist the infinitesimal local symmetries with a number of local parameters $\epsilon ^a$ equal to the number of the primary first class constraints
%${\stackrel{(N-1)}{\epsilon}}$-type
\begin{eqnarray}\label{01}
\delta q^B=\epsilon^a R_a^{(0)B}+\dot\epsilon^a R_a^{(1)B}+\ddot\epsilon^a R_a^{(2)B}+\ldots+
{\stackrel{(N-1)}{\epsilon}}{}^aR_a^{(N-1)B}.
\end{eqnarray}
Here $q^B$ is the set of configuration space variables,
${\stackrel{(k)}{\epsilon}}{}^a$ $\equiv$ $\frac{d^k\epsilon^a}{d\tau^k}$, and the set $R_a^{(k)B}(q, \dot q, \ldots)$ represents generator of the symmetry. In some particular models, the generators can be found in terms of constraints. For example, the relativistic particle Lagrangian $L$$=$$\sqrt{(\dot x^\mu)^2}$ implies the constraint $T$$\equiv$$\frac12(p^2-1)$, and the local symmetry $\delta x^\mu$$=$$\epsilon\frac{\dot x^\mu}{\sqrt{\dot x^2}}$. The latter can be rewritten as follows
\begin{eqnarray}\label{02}
\delta x^\mu=
\left.\epsilon\{x^\mu, T\}
\right|_{p_\mu\rightarrow\frac{\partial L}{\partial\dot x^\mu}},
\end{eqnarray}
where $\{{},{}\}$ is the Poisson bracket, and the symbol $|$ implies the indicated substitution. The equation (\ref{02}) states that the gauge generator is the Lagrangian counterpart of the canonical transformation generated by the constraint on a phase space. It seems to be interesting to find a proper generalization of the recipe given by Eq. (\ref{02}) on a general case. Since the Hamiltonian constraints can be found in the course of Dirac procedure, it would give a regular method for obtaining the symmetries.

General analysis of symmetry structure (classification and proof on existence of irreducible complete set of gauge generators) can be found in [13, 14].
In the works [5] it have been observed that symmetries of the extended Hamiltonian with first class constraints can be written in closed form. This observation was used in [4] to formulate the procedure for restoration of symmetries of the Hamiltonian action. While the algorithm suggested is relatively simple, some of its points remain unclarified. In particular, the completeness and irreducibility of the symmetries of the complete Hamiltonian were not demonstrated so far [13]. The Lagrangian symmetries have not been discussed. Analysis of a general case (when both first and second class constraints are present) turns out to be a much more complicated issue (see the second model of the Section 5 for the example). For the case, various procedures has been suggested and discussed in the works [6-10, 14-16].

We show that namely in the extended Lagrangian formalism the problem has a simple solution.
Complete irreducible set of local symmetries of $\tilde L$ will be presented in closed form through the first class constraints of the initial formulation, see Eq. (\ref{30}), (\ref{30.1}).
Moreover, all the initial variables $q^A$ transform according to Eq. (\ref{02}).

Another closely related issue is known as the Dirac conjecture [1]: does all the higher stage constraints generate the local symmetries? Affirmative answer on the question has been obtained by various groups [3, 8] in the extended Hamiltonian framework. Our result (\ref{30}) can be considered as another proof of the Dirac conjecture, now in the Lagrangian framework.

The work is organized as follows. With the aim to fix our notations, we outline in Section 2 the
Hamiltonization procedure for an arbitrary singular Lagrangian theory.
In Section 3 we formulate pure algebraic recipe for construction of the extended Lagrangian. All the higher-stage
constraints of $L$ appear as the second stage constraints in the formulation with $\tilde L$. Besides, we demonstrate that $\tilde L$ is a theory
with at most third-stage constraints. Then it is proved that $\tilde L$ and $L$ are
equivalent. It means, that an arbitrary theory can be reformulated as a theory with at most third-stage constraints\footnote{Popular physical theories usually do not involve more than third-stage constraints. Our result
can be considered as an explanation of this fact.}.
Since the original and the reconstructed formulations are equivalent,
it is matter of convenience to use one or another of them for description of the theory under
investigation.
In Section 4 we demonstrate one of advantages of the extended Lagrangian presenting its complete
irreducible set of local symmetry generators in terms of constraints. The procedure is illustrated on various examples in the Section 5.

\section{Dirac-Bergmann procedure for singular Lagrangian theory}
Let $L(q^A, \dot q^B)$ be Lagrangian of the singular theory:
$rank\frac{\partial^2 L}{\partial\dot q^A\partial\dot q^B}=[i]<[A]$, defined on
configuration space $q^A, A=1, 2, \ldots , [A]$. From the beginning, it is convenient to rearrange the  initial
variables in such a way
that the rank minor is placed in the upper left corner of the matrix
$\frac{\partial^2 L}{\partial\dot q^A\partial\dot q^B}$.
Then one has $q^A=(q^i, q^\alpha)$,
$i=1, 2, \ldots , [i]$, ~
$\alpha=1, 2, \ldots , [\alpha]=[A]-[i]$, where
$\det\frac{\partial^2 L}{\partial\dot q^i\partial\dot q^j}\ne 0$.

Let us construct the Hamiltonian formulation for the theory. To fix our notations, we carry out the
Hamiltonization procedure in some details. One introduces conjugate
momenta according to the equations
$p_i$ $=$ $\frac{\partial L}{\partial\dot q^i}$, $p_\alpha$ $=$ $\frac{\partial L}{\partial\dot q^\alpha}$.
They are considered as algebraic equations for determining velocities $\dot q^A$.
According to the rank condition, the first $[i]$ equations
can be resolved with respect to $\dot q^i$, let us denote the solution as
\begin{eqnarray}\label{2}
\dot q^i=v^i(q^A, p_j, \dot q^\alpha).
\end{eqnarray}
It can be substituted into remaining $[\alpha]$ equations for the momenta. By construction, the
resulting expressions do not depend on $\dot q^A$ and are called primary constraints
$\Phi_\alpha(q, p)$ of the Hamiltonian formulation. One finds
\begin{eqnarray}\label{3}
\Phi_\alpha\equiv p_\alpha-f_\alpha(q^A, p_j)=0,
\end{eqnarray}
where
\begin{eqnarray}\label{4}
f_\alpha(q^A, p_j)\equiv\left.\frac{\partial L}{\partial\dot q^\alpha}
\right|_{\dot q^i=v^i(q^A, p_j, \dot q^\alpha)}.
\end{eqnarray}
The original equations for the momenta are thus equivalent to the system (\ref{2}), (\ref{3}).
By construction, there are the identities
\begin{eqnarray}\label{1}
\left.\frac{\partial L(q, \dot q)}{\partial\dot q^i}\right|_{\dot q^i\rightarrow v^i(q^A, p_j, \dot q^\alpha)}\equiv p_i, \qquad
\left.v^i(q^A, p_j, \dot q^\alpha)\right|_{p_j\rightarrow\frac{\partial L}{\partial\dot q^j}}\equiv\dot q^i.
\end{eqnarray}
Next step of the Hamiltonian procedure is to introduce an extended phase space parameterized by the
coordinates $q^A, p_A, v^\alpha$, and to define the complete Hamiltonian $H$ according to the rule
\begin{eqnarray}\label{5}
H(q^A, p_A, v^\alpha)=H_0(q^A, p_j)+v^\alpha\Phi_\alpha(q^A, p_B),
\end{eqnarray}
where
\begin{eqnarray}\label{6}
H_0=\left.(p_i\dot q^i-L+ \dot q^\alpha\frac{\partial L}{\partial \dot q^\alpha})\right|
_{\dot q^i\rightarrow v^i(q^A, p_j, \dot q^\alpha)}.
\end{eqnarray}
By construction it does not contain the quantities $\dot q^\alpha$ and $p_\alpha$.
The Hamiltonian equations
\begin{eqnarray}\label{7}
\dot q^A=\{q^A, H\}, \qquad \dot p_A=\{p_A, H\}, \qquad
\Phi_\alpha(q^A, p_B)=0,
\end{eqnarray}
are equivalent to the Lagrangian equations following from $L$, see [3]. Here $\{ , \}$ denotes
the Poisson bracket.

From Eq. (\ref{7}) it follows that all the solutions are confined to lie on a surface of the extended phase space
defined by the algebraic equations $\Phi_\alpha=0$. It may happen, that
the system (\ref{7}) contains in reality more then $[\alpha]$ algebraic equations. Actually, derivative of the
primary constraints with respect to time implies, as algebraic consequences of the system (\ref{7}),
the so called second stage equations: $\{\Phi_\alpha, H\}$ $\equiv$
$\{\Phi_{\alpha}, \Phi_\beta\}v^\beta+\{\Phi_{\alpha}, H_0\}$ $=$ $0$.
They can be added to Eq. (\ref{7}), which gives an equivalent system.
Let on-shell one has $rank\{\Phi_{\alpha}, \Phi_\beta\}=[\alpha ']\leq [\alpha]$. Then
$[\alpha ']$
equations of the second-stage system
can be used to represent some $v^{\alpha '}$ through other variables. It can be substituted into the remaining
$[\alpha '']\equiv [\alpha]-[\alpha ']$ equations, the resulting expressions do not contain $v^\alpha$ at all.
Thus the second-stage system can be presented in the equivalent form
\begin{eqnarray}\label{7.0}
v^{\alpha '}=v^{\alpha '}(q^A, p_j, v^{\alpha ''}), \qquad T_{\alpha ''}(q^A, p_j)=0.
\end{eqnarray}
Functionally independent equations among $T_{\alpha ''}=0$, if any, represent secondary Dirac constraints.
Thus all the
solutions of the system (\ref{7}) are confined to the surface defined by $\Phi_\alpha=0$ and by the
equations (\ref{7.0}).

The secondary constraints may
imply third-stage constraints, and so on. We suppose that the theory has
constraints up to $N$-th stage, $N\ge 2$. The complete set of higher stage constraints is denoted by $T_a(q^A, p_j)=0$.
Then the complete constraint system is
$G_I\equiv(\Phi_\alpha, T_a)$. All the solutions of Eq. (\ref{7}) are confined to the
surface defined by the equations $\Phi_\alpha=0$ as well as by\footnote{It is known
[3], that the procedure reveals all the algebraic equations presented in the system (\ref{7}). Besides, surface
of solutions of Eq. (\ref{7}) coincides with the surface $\Phi_\alpha=0$, $\{ G_I, H\}=0$.}
\begin{eqnarray}\label{7.1}
\{ G_I, H\}=0.
\end{eqnarray}
By construction, after substitution of the velocities $v^\alpha$ determined in the course of Dirac procedure, the equations (\ref{7.1}) vanish
on the complete constraint surface $G_J$$=$$0$.

Suppose that $\{G_I, G_J\}=\triangle_{IJ}(q^A, p_j)$, where
$\left. rank\triangle_{IJ}\right|_{G_I=0}=[I_2]<[I]$. It means that both first and second class constraints are presented in the formulation.
It will be convenient to separate them. According to the rank condition, there exist $[I_1]={I}-[I_2]$
independent null-vectors $\vec K_{I_1}$ of the matrix $\triangle$ on the surface $G_I=0$, with the components
$K_{I_1}{}^J(q^A, p_j)$. Then the bracket of constraints $G_{I_1}\equiv K_{I_1}{}^JG_J$ with any $G_I$ vanishes, hence the constraints
$G_{I_1}$ represent the first class subset. One chooses the vectors $\vec K_{I_2}(q^A, p_j)$ to complete
$K_{I_1}$ up to a basis of $[I]$-dimensional vector space. By construction, the matrix
\begin{eqnarray}\label{7.2}
K_{I}{}^J\equiv
\left(K_{I_1}{}^J\atop  K_{I_2}{}^J\right),
\end{eqnarray}
is invertible. Let us denote
$\tilde G_I$$\equiv$$(\tilde G_{I_1}$, $\tilde G_{I_2})$, where
$\tilde G_{I_1}\equiv K_{I_1}{}^JG_J$, $\tilde G_{I_2}\equiv K_{I_2}{}^JG_J$.
The system $\tilde G_I$
is equivalent to the initial system of constraints $G_I$. The constraints $\tilde G_{I_2}$ form the second class
subset of the complete set $\tilde G_I$. In an arbitrary theory, the constraints obey the following Poisson bracket algebra:
\begin{eqnarray}\label{8}
\{\tilde G_I, \tilde G_J\}=\triangle_{IJ}(q^A, p_B), \qquad \quad \qquad \qquad \quad \cr
\{\tilde G_{I_1}, G_J\}=c_{I_1 J}{}^K(q^A, p_B)G_K, \quad \{\tilde G_{I_1}, H_0\}=b_{I_1}{}^J(q^A, p_B)G_J, \cr
\{\tilde G_{I_2}, \tilde G_{J_2}\}=\triangle_{I_2 J_2}(q^A, p_B), \qquad \qquad \quad \qquad
\end{eqnarray}
where
\begin{eqnarray}\label{8.1}
\left. rank\triangle_{IJ}\right|_{G_I=0}=[I_2], \qquad
\left. \det\triangle_{I_2 J_2}\right|_{G_I=0}\ne 0.
\end{eqnarray}
The extended Hamiltonian is defined as follows
\begin{eqnarray}\label{8.1}
H_{ext}(q^A, p_A, v^\alpha, \lambda^a)=H_0(q^A, p_j)+v^\alpha\Phi_\alpha(q^A, p_j, p_\alpha)+
\lambda^aT_a(q^A, p_j),
\end{eqnarray}
As it was mentioned in the introduction, $H_{ext}$ cannot be generally obtained as the complete Hamiltonian of some Lagrangian. It can be seen as follows. In the Dirac-Bergmann procedure, the total Hamiltonian is uniquely defined by Eqs. (\ref{5}), (\ref{6}). Consider the particular case of higher stage constraints $T_a$ of the form $p_a$: $T_a$ $=$ $p_a$ $-$ $t_a(q^A, p')$. Then it is clear that Eq. (\ref{8.1}) does not have the desired form (\ref{5}), since $H_0$ from (\ref{8.1}) generally depends on $p_a$.

\section{Formalism of extended Lagrangian}
Starting from the theory described above, we
construct here the equivalent Lagrangian $\tilde L(q^A, \dot q^A, s^a)$ defined on the
configuration space with the coordinates $q^A, s^a$, where $s^a$ states for auxiliary variables. By construction, it will generate the Hamiltonian of the form
$H_0+s^aT_a$, as well as the primary constraints $\Phi_\alpha=0$, $\pi_a=0$,
where $\pi_a$ represent
conjugate momenta\footnote{Let us stress once again, that in our formulation the
variables $s^a$ represent a part of the configuration-space variables.} for $s^a$. Due to the special form of Hamiltonian, preservation in time of the primary constraints
$\pi_a$$=$$0$
implies that all the higher stage constraints $T_a$ of the original formulation appear as the secondary constraints of $\tilde L$: $\dot\pi_a$$=$$\{\pi_a, H_0+s^aT_a\}$$=$$-T_a$$=$$0$.

To construct the extended Lagrangian for $L$, one introduces the following equations for
the variables\footnote{As it will be shown below, Eq. (\ref{9}) represents a solution of  the equation $\tilde p_j$$=$$\frac{\partial\tilde L}{\partial\dot q^j}$
defining the conjugate momenta $\tilde p_j$ of the extended formulation.}
$q^A$, $\tilde p_j$, $s^a$:
\begin{eqnarray}\label{9}
\dot q^i-v^i(q^A, \tilde p_j, \dot q^\alpha)-s^a\frac{\partial T_a(q^A, \tilde p_j)}{\partial \tilde p_i}=0.
\end{eqnarray}
Here the functions $v^i(q^A, \tilde p_j, \dot q^\alpha)$, ~ $T_a(q^A, \tilde p_j)$
are taken from the initial formulation.
The equations can be resolved algebraically with respect to $\tilde p_i$ in a vicinity  of the
point $s^a=0$. Actually, Eq. (\ref{9}) with $s^a=0$ coincides with Eq. (\ref{2}) of the initial formulation,
the latter can be resolved, see Eq. (\ref{1}). Hence
$\det\frac{\partial (Eq.(\ref{9}))^i}{\partial\tilde p_j}\ne 0$ at the
point $s^a=0$. Then the same is true in some vicinity of this point, and Eq. (\ref{9})
thus can be resolved. Let us denote the solution as
\begin{eqnarray}\label{10}
\tilde p_i=\omega_i(q^A, \dot q^A, s^a).
\end{eqnarray}
By construction, there are the identities
\begin{eqnarray}\label{11}
\left.\omega_i(q, \dot q, s)\right|_{\dot q^i\rightarrow
v^i(q^A, \tilde p_j, \dot q^\alpha)+s^a\frac{\partial T_a(q^A, \tilde p_j)}{\partial\tilde p_i}}\equiv\tilde p_i,
\end{eqnarray}
\begin{eqnarray}\label{11.-1}
\left.\left(v^i(q^A, \tilde p_j, \dot q^\alpha)+s^a\frac{\partial T_a(q^A, \tilde p_j)}{\partial\tilde p_i}\right)
\right|_{\omega_i(q, \dot q, s)}\equiv\dot q^i.
\end{eqnarray}
Besides, the function $\omega$ has the property
\begin{eqnarray}\label{12}
\left.\omega_i(q^A, \dot q^A, s^a)\right|_{s^a=0}=\frac{\partial L}{\partial\dot q^i}.
\end{eqnarray}
Now, the extended Lagrangian for $L$ is defined according to the expression
\begin{eqnarray}\label{13}
\tilde L(q^A, \dot q^A, s^a)=L(q^A, v^i(q^A, \omega_j, \dot q^\alpha), \dot q^\alpha)+ \cr
\omega_i(\dot q^i-v^i(q^A, \omega_j, \dot q^\alpha))-s^aT_a(q^A, \omega_j), \quad
\end{eqnarray}
where the functions $v^i, \omega_i$ are given by Eqs. (\ref{2}), (\ref{10}).
As compare with the initial Lagrangian, $\tilde L$ involves the new variables $s^a$, in a number
equal to the number of higher stage constraints $T_a$. Let us enumerate some properties
of $\tilde L$
\begin{eqnarray}\label{14}
\tilde L(s^a=0)=L,
\end{eqnarray}
\begin{eqnarray}\label{15}
\left.\frac{\partial\tilde L}{\partial\omega_i}
\right|_{\omega(q, \dot q, s)}=0,
\end{eqnarray}
\begin{eqnarray}\label{16}
\frac{\partial\tilde L}{\partial\dot q^\alpha}=
\left.\frac{\partial L(q^A, v^i, \dot q^\alpha)}{\partial\dot q^\alpha}
\right|_{v^i(q, \omega, \dot q^\alpha)}=
f_\alpha(q^A, \omega_j(q, \dot q, s)).
\end{eqnarray}
Eq. (\ref{14}) follows from Eqs. (\ref{12}), (\ref{1}). Eq. (\ref{15}) is a consequence of the identities (\ref{1}),
(\ref{11}). Eq. (\ref{15}) will be crucial for discussion of local symmetries
in the next section. At last, Eq. (\ref{16}) is a consequence of Eqs. (\ref{15}), (\ref{1}).

Following to the standard prescription [3, 4], let us construct the Hamiltonian formulation
for $\tilde L$. By using of Eqs. (\ref{15}), (\ref{16}), one finds the conjugate momenta $\tilde p_A$, $\pi_a$ for $q^A, s^a$
\begin{eqnarray}\label{17.1}
\tilde p_i=\frac{\partial\tilde L}{\partial\dot q^i}=\omega_i(q^A, \dot q^A, s^a),
\end{eqnarray}
\begin{eqnarray}\label{17}
\tilde p_\alpha=\frac{\partial\tilde L}{\partial\dot q^\alpha}=f_\alpha(q^A, \omega_j),
\cr
\pi_a=\frac{\partial\tilde L}{\partial\dot s^a}=0. \qquad \qquad
\end{eqnarray}
The equation (\ref{17.1}) can be resolved with respect to the velocities $\dot q^i$. According to the identity (\ref{11}), the solution is just given by our basic equation (\ref{9}). Taking this into account, the system (\ref{17.1}), (\ref{17}) is equivalent to the following one
\begin{eqnarray}\label{18.1}
\dot q^i=v^i(q^A, \tilde p_j, \dot q^\alpha)+s^a\frac{\partial T_a(q^A, \tilde p_j)}{\partial\tilde p_i},
\end{eqnarray}
\begin{eqnarray}\label{18}
\tilde p_\alpha-f_\alpha(q^A, \tilde p_j)=0,
\end{eqnarray}
\begin{eqnarray}\label{18.2}
\pi_a=0.
\end{eqnarray}
So, in the extended formulation there are presented the primary constraints (\ref{18}) of the initial formulation. Besides, there are the trivial constraints (\ref{18.2}) in a number equal to the number of all the higher stage constraints of the initial formulation.

Using the definition (\ref{6}), one obtains the Hamiltonian
$\tilde H_0$ $=$ $H_0+s^aT_a$, then the complete Hamiltonian for $\tilde L$ is given by the expression
\begin{eqnarray}\label{19}
\tilde H%(q^A, \tilde p_A, s^a, \pi_a, v^\alpha, v^a)
=H_0(q^A, \tilde p_j)+
s^aT_a(q^A, \tilde p_j)
+v^\alpha\Phi_\alpha(q^A, \tilde p_B)+v^a\pi_a.
\end{eqnarray}
Here $v^\alpha, v^a$ are the primarily un expressible velocities of $\tilde L$.
Note that, if one discards the constraints $\pi_a=0$, $\tilde H$ coincides with the extended Hamiltonian for $L$ after
identification of the configuration space variables $s^a$ with the Lagrangian
multipliers for higher stage constraints of the original formulation.

Further, preservation in time of the
primary constraints: $\dot\pi_a$$=$$\{\pi_a,$ $H_0+s^aT_a\}$$=$$-T_a$$=$$0$ implies the equations $T_a=0$. Hence all the higher stage constraints of the initial
formulation appear now as the secondary constraints. Preservation in time of the primary constraints $\Phi_\alpha$
leads to the equations $\{\Phi_\alpha, \tilde H\}$ $=$
$\{\Phi_\alpha, H_0\}$ $+$ $\{\Phi_\alpha, \Phi_\beta\}v^\beta$ $+$ $\{\Phi_\alpha, T_b\}s^b$ $=$ $0$.
In turn, preservation of the secondary constraints $T_a$ leads to the similar equations
$\{T_a, \tilde H\}$ $=$
$\{T_a, H_0\}$ $+$ $\{T_a, \Phi_\beta\}v^\beta$ $+$ $\{T_a, T_b\}s^b$ $=$ $0$. To continue the analysis, it is
convenient to unify them as follows:
\begin{eqnarray}\label{20}
\{G_I, H_0\}+\{G_I, G_J\}S^J=0.
\end{eqnarray}
Here $G_I$ are all the constraints of the initial formulation and it was denoted $S^J\equiv(v^\alpha, s^a)$.
Using the matrix (\ref{7.2}), the system (\ref{20}) can be rewritten in the equivalent form
\begin{eqnarray}\label{21}
\{\tilde G_{I_1}, H_0\}+O(G_I)=0,
\end{eqnarray}
\begin{eqnarray}\label{22}
\{\tilde G_{I_2}, H_0\}+\{\tilde G_{I_2}, G_J\}S^J=O(G_I).
\end{eqnarray}
Eq. (\ref{21}) does not contain any new information, since the first class constraints commute with the Hamiltonian,
see Eq. (\ref{8}). Let us analyze the system (\ref{22}). First, one notes that due to the rank condition
$\left.rank\{\tilde G_{I_2}, G_J\}\right|_{G_I}$ $=$ $[I_2]=max$, exactly $[I_2]$ variables among $S^I$ can be determined from
the system. According to the Dirac prescription, one needs to determine the maximal number of the multipliers $v^\alpha$. To make this,
let us restore $v$-dependence in Eq. (\ref{22}):
$\{\tilde G_{I_2}, \Phi_\alpha\}v^\alpha$ $+$ $\{\tilde G_{I_2}, H_0\}+\{\tilde G_{I_2}, T_b\}s^b$ $=$ $0$.
Since the matrix $\{\tilde G_{I_2}, \Phi_\alpha\}$ is the same as in the initial formulation, from these equations one
determines some group of variables $v^{\alpha_2}$ through the remaining variables $v^{\alpha_1}$, where $[\alpha_2]$
is the number of primary second-class constraints among $\Phi_\alpha$. After substitution of the result into the
remaining equations of the system (\ref{22}), the latter acquires the form
\begin{eqnarray}\label{23}
v^{\alpha_2}=v^{\alpha_2}(q, \tilde p, s^a, v^{\alpha_1}), \qquad
Q_{a_2 b}(q, \tilde p)s^b+P_{a_2}(q, \tilde p)=0,
\end{eqnarray}
where $[a_2]$ is the number of higher-stage second class constraints of the initial theory.
It must be $P\approx 0$, since for $s^b=0$ the system (\ref{22}) is a subsystem of
(\ref{7.1}), but the latter vanish after substitution of the multipliers determined during the procedure,
see discussion after Eq. (\ref{7.1}). Besides, one notes that $rank Q=[a_2]=max$. Actually, suppose that
$rank Q=[a']<[a_2]$. Then from Eq. (\ref{22}) only $[\alpha_2]+[a']<[I_2]$
variables among $S^I$ can be determined, in contradiction with the conclusion made before. In resume, the
system (\ref{20}) for determining the second-stage and third-stage constraints and multipliers is
equivalent to
\begin{eqnarray}\label{24}
v^{\alpha_2}=v^{\alpha_2}(q, \tilde p, s^{a_1}, v^{\alpha_1}),
\end{eqnarray}
\begin{eqnarray}\label{25}
s^{a_2}=\tilde Q^{a_2}{}_{b_1}(q, \tilde p)s^{b_1},
\end{eqnarray}
with some matrix $\tilde Q$.
Conservation in time of the constraints (\ref{25}) leads to the equations for
determining the multipliers
\begin{eqnarray}\label{25.1}
v^{a_2}=\{ Q^{a_2}{}_{b_1}(q, \tilde p)s^{b_1}, \tilde H\}.
\end{eqnarray}
Since there are no new constraints,
the Dirac procedure for $\tilde L$ stops on this stage. All the constraints of the theory have been revealed
after completing the third stage.

Now we are ready to compare the theories $\tilde L$ and $L$. Dynamics of the theory $\tilde L$ is governed by the
Hamiltonian equations
\begin{eqnarray}\label{26}
\dot q^A=\{q^A, H\}+s^a\{q^A, T_a\}, \qquad \dot{\tilde p}_A=\{\tilde p_A, H\}+s^a\{\tilde p_A, T_a\}, \cr
\dot s^a=v^a, \qquad \qquad \qquad \qquad \qquad \dot\pi_a=0, \qquad \qquad \qquad \qquad \quad
\end{eqnarray}
as well as by the constraints
\begin{eqnarray}\label{27}
\Phi_\alpha=0, \qquad T_a=0,
\end{eqnarray}
\begin{eqnarray}\label{28}
\pi_{a_1}=0,
\end{eqnarray}
\begin{eqnarray}\label{29}
\pi_{a_2}=0, \qquad s^{a_2}=Q^{a_2}{}_{b_1}(q, \tilde p)s^{b_1}.
\end{eqnarray}
Here $H$ is the complete Hamiltonian of the initial theory (\ref{5}), and the Poisson bracket is defined
on the phase space $q^A, s^a, \tilde p_A, \pi_a$. The constraints $\pi_{a_1}=0$ can be replaced by the combinations
$\pi_{a_1}+\pi_{a_2}Q^{a_2}{}_{a_1}(q, \tilde p)=0$, the latter represent first class subset.
Let us make partial fixation of a gauge by imposing the equations $s^{a_1}=0$ as a gauge conditions for the
subset. Then $(s^a, \pi_a)$-sector of the theory disappears, whereas the equations (\ref{26}), (\ref{27}) coincide
exactly with those of the initial
theory\footnote{In more rigorous treatment,
one writes Dirac bracket corresponding to the equations $\pi_{a_1}-\pi_{a_2}Q^{a_2}{}_{a_1}=0$, $s^{a_1}=0$,
and to the second class constraints (\ref{29}). After that, the equations used in construction of the Dirac bracket
can be used as strong equalities. For the case, they reduce to the equations $s^a=0, \pi_a=0$. For the remaining phase-space
variables $q^A, p_A$, the Dirac bracket coincides with the Poisson one.} $L$.
Let us reminded that $\tilde L$
has been constructed in some vicinity of the point $s^a=0$. The gauge $s^{a_1}=0$ implies $s^a=0$ due to the homogeneity
of Eq. (\ref{25}). It guarantees a self consistency of the construction. Thus $L$ represents one of the
gauges [3] for $\tilde L$, which proves an equivalence of the two formulations.

Using Eqs. (\ref{11}) (\ref{11.-1}), the extended Lagrangian (\ref{13}) can be rewritten in the equivalent form
\begin{eqnarray}\label{29.10}
\tilde L(q^A, \dot q^A, s^a)=L(q^A, \dot q^i-
s^a\frac{\partial T_a(q^a, \omega_i)}{\partial\omega_i}, \dot q^\alpha)+ \cr
s^a(\omega_i\frac{\partial T_a(q^a, \omega_i)}{\partial\omega_i}-T_a(q^A, \omega_i))
\end{eqnarray}
Modulo to the extra term represented by the second line in Eq. (\ref{29.10}), $\tilde L$ is obtained from $L$ replacing the derivative $\dot q^i$ by the quantity similar to the covariant derivative
\begin{eqnarray}\label{29.11}
\partial_\tau q^i \longrightarrow D_\tau q^i=\partial_\tau q^i-
s^a\frac{\partial T_a(q^a, \omega_i)}{\partial\omega_i}.
\end{eqnarray}
The second line in Eq.(\ref{29.10}) disappears when the higher stage constraints are homogeneous on momenta. For example, for the constraints of the form\footnote{It is known that any first class system acquires this form in special canonical variables [3].} $T_ a$ $=$ $p_a$, where $p_a$ is a part of the momenta $p_i$ $=$ $(p_a, p'_i)$, the extended action acquires the form
\begin{eqnarray}\label{29.12}
\tilde L=L(q^A, \dot q^a-s^a, \dot q'^i, \dot q^\alpha).
\end{eqnarray}
For the case $T_a$ $=$ $h_a{}^i(q)p_i$ the extended Lagrangian is
\begin{eqnarray}\label{29.13}
\tilde L=L(q^A, \dot q^i-s^ah_a{}^i, \dot q^\alpha).
\end{eqnarray}
In both cases, it can be shown that $\tilde L$ is invariant under the local transformations with the transformation law for $s^a$ being proportional to $\dot\epsilon^a$. So, at least for these particular examples, $s^a$ can be identified with a gauge field supplying the local symmetry. It leads to the suggestion that in the passage from $\L$ to $\tilde L$
the local symmetries with higher order derivatives of the local parameters decompose into a sum of the gauge symmetries (with at most one derivative acting on the parameters). We confirm this statement in the next section.

\section{Local symmetries of the extended Lagrangian. Dirac conjecture.}

Since the initial Lagrangian is a gauge for the extended one, the physical system under consideration can
be equally analyzed using the extended Lagrangian. Higher stage constraints $T_a$ of $L$ turn out to be the second stage constraints of $\tilde L$. They enter into the expressions for $\tilde L$ and $\tilde H$ in the manifest form, see Eqs. (\ref{13}),(\ref{19}).
Here we demonstrate one of consequences of this
property: all the infinitesimal local symmetries of $\tilde L$ are the gauge symmetries and can be found in closed form in terms of the first class constraints.

According to the analysis made in the previous section, the primary constraints of the extended formulation are
$\Phi_\alpha=0$, $\pi_a=0$. Among $\Phi_\alpha=0$ there are presented first class constraints, in a
number equal to the number of primary first class constraints of $L$. Among
$\pi_a=0$, we have found the first class constraints $\pi_{a_1}-\pi_{a_2}Q^{a_2}{}_{a_1}(q, p)=0$, in
a number equal to the number of all the higher-stage first class constraints of $L$.
Thus the number of primary first class constraints of $\tilde L$ coincide with the number $[I_1]$
of all the first class constraints of $L$. Hence one expects $[I_1]$ local symmetries presented in the
formulation $\tilde L$.
Now we demonstrate that they are:
\begin{eqnarray}\label{30}
\delta_{I_1} q^A=\epsilon^{I_1}\left.\{q^A, \tilde G_{I_1}(q^A, \tilde p_B)\}
\right|_{\tilde p_i\rightarrow\frac{\partial\tilde L}{\partial\dot q^i}},
\end{eqnarray}
\begin{eqnarray}\label{30.1}
\delta_{I_1} s^a=
\left.\left[\dot\epsilon^{I_1}K_{I_1}{}^a+\epsilon^{I_1}\left( b_{I_1}{}^a+s^bc_{I_1 b}{}^a+
\dot q^\beta c_{I_1 \beta}{}^a\right)\right]
\right|_{\tilde p_i\rightarrow\frac{\partial\tilde L}{\partial\dot q^i}}.
\end{eqnarray}
Here $\epsilon^{I_1}(\tau)$, $I_1=1, 2, \ldots , [I_1]$ are the local parameters, and $K$ is the conversion
matrix, see Eq. (\ref{7.2}).

According to Eq. (\ref{30.1}) variation of some $s^a$ involve derivative of parameters. Hence they can be identified with a gauge fields for the symmetry. At this point,
it is instructive to discuss what happen with local symmetries on the passage from $L$ to $\tilde L$.
Appearance of some $N$-th stage first-class constraint in the Hamiltonian formulation for $L$ implies [15], that $L$ has the local symmetry of ${\stackrel{(N-1)}{\epsilon}}$-type (\ref{01}).
Replacing $L$ with $\tilde L$, one arrives at the formulation with the secondary first class constraints and the corresponding $\dot\epsilon$-type symmetries (\ref{30}). That is  the symmetry (\ref{01}) of $L$ "decomposes" into $N$ gauge symmetries of $\tilde L$.

According to Eq. (\ref{30}), transformations of the original variables $q^A$ are generated by all the first class constraints of initial formulation. This result can be considered as a proof of the Dirac conjecture.

We now show that variation of $\tilde L$ under the transformation (\ref{30}) is proportional to the higher stage constraints $T_a$. So, it can be canceled by appropriate variation of $s^a$, the latter turns out to be given by Eq. (\ref{30.1}).
In the subsequent computations we omit all the total derivative terms. Besides,
the notation $\left. A\right|$ implies the substitution indicated in Eqs. (\ref{30}), (\ref{30.1}).

To make a proof, it is convenient to represent the extended Lagrangian (\ref{13}) in terms of the initial
Hamiltonian $H_0$, instead of the initial Lagrangian $L$. Using Eq. (\ref{6}) one writes
\begin{eqnarray}\label{31}
\tilde L(q^A, \dot q^A, s^a)=
\omega_i\dot q^i+f_\alpha(q^A, \omega_j)\dot q^\alpha-
H_0(q^A, \omega_j)-s^aT_a(q^A, \omega_j),
\end{eqnarray}
where the functions $\omega_i(q, \dot q, s)$, $f_\alpha(q, \omega)$ are defined by Eqs. (\ref{10}), (\ref{4}). According to the identity (\ref{15}), variation of $\tilde L$ with respect to $\omega_i$ does not give any contribution. Taking this into account, variation of Eq. (\ref{31}) under the transformation (\ref{30}) can be
written in the form
\begin{eqnarray}\label{32}
\delta\tilde L=-\dot\omega_i(q, \dot q, s)\left.\frac{\partial\tilde G_{I_1}}{\partial \tilde p_i}\right|\epsilon^{I_1}
-\dot f_\alpha(q, \omega(q, \dot q, s)\left.\frac{\partial\tilde G_{I_1}}{\partial \tilde p_\alpha}\right|\epsilon^{I_1}
\qquad \qquad \quad  ~ \cr
-\left. \left(\frac{\partial H_0(q^A, \tilde p_j)}{\partial q^A}+
\dot q^\alpha\frac{\partial\Phi_\alpha(q^A, \tilde p_B)}{\partial q^A}+
s^a\frac{\partial T_a(q^A, \tilde p_j)}{\partial q^A}\right)\right|\left.\{q^A, \tilde G_{I_1}\}\right|\epsilon^{I_1} \cr
-\delta_{I_1}s^aT_a(q^A, \omega_j).\qquad \qquad \qquad \qquad \qquad \qquad \qquad \qquad \qquad \qquad \quad
\end{eqnarray}
To see that $\delta\tilde L$ is the total derivative, we add the following zero
\begin{eqnarray}\label{33}
0\equiv\left.\left[\left.\frac{\partial\tilde L}{\partial\omega_i}
\right|_{\omega_i}\{\tilde p_i, \tilde G_{I_1}\}\right.\right.
\qquad \qquad \qquad \qquad \qquad \qquad \qquad \quad ~\cr
\left.\left.
-\left(\frac{\partial H_0}{\partial \tilde p_\beta}+
\dot q^\alpha\frac{\partial\Phi_\alpha}{\partial \tilde p_\beta}+
s^a\frac{\partial T_a}{\partial \tilde p_\beta}\right)\{\tilde p_\beta, \tilde G_{I_1}\}+
\dot q^\alpha\{\tilde p_\alpha, \tilde G_{I_1}\}\right]\right|\epsilon^{I_1},
\end{eqnarray}
to the r.h.s. of Eq. (\ref{32}). It leads to the expression
\begin{eqnarray}\label{34}
\delta\tilde L=
\left.\left[\dot\epsilon^{I_1}\tilde G_{I_1}-\epsilon^{I_1}\left(\{H_0, \tilde G_{I_1}\}+
\dot q^\alpha\{\Phi_\alpha, \tilde G_{I_1}\}+s^a\{T_a, \tilde G_{I_1}\}\right)\right]\right| \cr
-\delta_{I_1}s^aT_a(q^A, \omega_j)= \qquad \qquad \qquad \qquad \qquad \qquad \qquad \qquad \qquad \cr
\left.\left[\dot\epsilon^{I_1}\tilde G_{I_1}+\epsilon^{I_1}\left(b_{I_1}{}^I+
\dot q^\alpha c_{I_1 \alpha}{}^I+s^bc_{I_1 b}{}^I\right)G_I\right]\right|-\delta_{I_1}s^aT_a(q^A, \omega_j),
\end{eqnarray}
where $b, c$ are coefficient functions  of the constraint algebra (\ref{8}).
Using the equalities $\left. G_{I}\right|=(0, ~  T_a(q^A, \omega_j))$,
$\left. \tilde G_{I_1}\right|=K_{I_1}{}^a T_a(q^A, \omega_j)$, one finally obtains
\begin{eqnarray}\label{35}
\delta\tilde L= \qquad \qquad \qquad \qquad \qquad \qquad \qquad \cr
\left.\left[\dot\epsilon^{I_1}K_{I_1}{}^a+\epsilon^{I_1}\left(b_{I_1}{}^a+
\dot q^\alpha c_{I_1 \alpha}{}^a+s^bc_{I_1 b}{}^a\right)-\delta_{I_1}s^a\right]\right|_
{p_i\rightarrow\omega_i}T_a.
\end{eqnarray}
Then the variation of $s^a$ given in Eq. (\ref{30}) implies $\delta\tilde L=div$, as it has been stated.

In the absence of second class constraints, Eqs. (\ref{30}), (\ref{30.1}) acquire the form
\begin{eqnarray}\label{30.-1}
\delta_I q^A=\epsilon^I\left.\{q^A, G_I(q^A, \tilde p_B)\}
\right|_{\tilde p_i\rightarrow\frac{\partial\tilde L}{\partial\dot q^i}}, \cr
\delta_I s^a=
\left.\left[\dot\epsilon^a\delta_{aI}+\epsilon^I\left( b_I{}^a+s^bc_{I b}{}^a+
\dot q^\beta c_{I \beta}{}^a\right)\right]
\right|_{\tilde p_i\rightarrow\frac{\partial\tilde L}{\partial\dot q^i}}.
\end{eqnarray}
They can be used to construct symmetries of the original Lagrangian. To this end, one notes that the extended Lagrangian coincides with the original one for $s^a=0$:
$\tilde L(q, 0)=L(q)$, see Eq. (\ref{14}). So the initial action will be invariant under any transformation
\begin{eqnarray}\label{25.1}
\delta q^A=\sum_{I_1}\left.\delta_I q^A\right|_{s=0},
\end{eqnarray}
which obeys to the system $\left.\delta s^a\right|_{s=0}=0$, that is
\begin{eqnarray}\label{25.2}
\dot\epsilon^IK_I{}^a+\epsilon^I\left( b_I{}^a+
\dot q^\beta c_{I \beta}{}^a\right)=0.
\end{eqnarray}
One has $[a]$ equations for $[\alpha]+[a]$ variables $\epsilon^I$. Similarly to Ref. [4], the equations can be solved by
pure algebraic methods, which give some $[a]$ of $\epsilon$ in terms of the remaining $\epsilon$ and their derivatives
of order less than $N$. It allows one to find $[\alpha]$ local symmetries of $L$. As it was already mentioned, the problem
here is to prove the completeness and the irreducibility of the set.

\section{Examples}
{\bf 1) Model with fourth-stage constraints.} Let us consider the Lagrangian
\begin{eqnarray}\label{E1}
L=\frac12(\dot x)^2+\xi (x)^2,
\end{eqnarray}
where $x^\mu(\tau), \xi(\tau)$ are configuration space variables, $\mu=0, 1, \ldots, n$,
$(x)^2\equiv\eta_{\mu\nu}x^\mu x^\nu$, $\eta_{\mu\nu}=(-, +, \ldots , +)$.

Denoting the conjugate momenta for
$x^\mu,  \xi$ as $p_\mu, p_{\xi}$, one obtains the complete Hamiltonian
\begin{eqnarray}\label{E2}
H_0=\frac12p^2-\xi (x)^2+v_{\xi}p_{\xi},
\end{eqnarray}
where $v_{\xi}$ is multiplier for the primary constraint $p_\xi=0$. The complete system of constraints turns out
to be
\begin{eqnarray}\label{E3}
\Phi_1\equiv p_{\xi}=0, \quad T_2\equiv x^2=0, \quad
T_3\equiv xp=0, \quad T_4\equiv p^2=0.
\end{eqnarray}
For the case, the variable $\xi$ plays the role of $q^\alpha$, while $x^\mu$ play the role of $q^i$ of the general formalism.

The constraints are first class
\begin{eqnarray}\label{E4}
\{G_I, G_J\}=c_{IJ}{}^{K}(q^A, p_j)G_K, \qquad \{G_I, H_0\}=b_{I}{}^J(q^A, p_j)G_J,
\end{eqnarray}
with non vanishing coefficient functions being
\begin{eqnarray}\label{E5}
c_{2 3}{}^2=-c_{3 2}{}^2=2, \qquad c_{2 4}{}^3=-c_{4 2}{}^3=4, \qquad c_{3 4}{}^4=-c_{4 3}{}^4=2; \cr
b_1{}^2=1, \qquad b_2{}^3=2, \qquad b_3{}^4=1, \qquad b_3{}^3=2\xi, \qquad b_4{}^3=4\xi.
\end{eqnarray}
For the present case, Eq. (\ref{9}) acquires the form
$\dot x^\mu$$-\tilde p^\mu$$-s^3x^\mu$$-2s^4\tilde p^\mu$ $=$ $0$, so
\begin{eqnarray}\label{E6}
\tilde p^\mu=\frac{1}{1+2s^4}(\dot x^\mu-s^3x^\mu).
\end{eqnarray}
The r.h.s. represents the function $\omega$ of the general formalism. Then the extended Lagrangian (\ref{29.10}) is given by
\begin{eqnarray}\label{E7}
\tilde L=\frac{1}{2(1+2s^4)}(\dot x^\mu-s^3x^\mu)^2+(\xi-s^2)(x^\mu)^2.
\end{eqnarray}
Using the equations (\ref{30.-1}), (\ref{E5}), its symmetries can be written immediately as follows
\begin{eqnarray}\label{E8}
\delta_1\xi=\epsilon^1, \qquad \delta_1s^2=\epsilon^1;\qquad \qquad \qquad \qquad \qquad \qquad \qquad \qquad \quad
\end{eqnarray}
\begin{eqnarray}\label{E9}
\delta_2s^2=\dot\epsilon^2+2\epsilon^2s^3,  \qquad \delta_2s^3=2\epsilon^2(1+2s^4);
\qquad \qquad \qquad \qquad ~
\end{eqnarray}
\begin{eqnarray}\label{E10}
\delta_3x^\mu=\epsilon^3 x^\mu, ~ \delta_3s^2=2\epsilon^3(\xi-s^2), ~
\delta_3s^3=\dot\epsilon^3, ~ \delta_3s^4=\epsilon^3(1+2s^4);
\end{eqnarray}
\begin{eqnarray}\label{E11}
\delta_4x^\mu=2\epsilon^4 \frac{\dot x^\mu-s^3x^\mu}{1+2s^4}, \quad \delta_4s^3=4\epsilon^4(\xi-s^2), \quad
\delta_4s^4=\dot\epsilon^4-2\epsilon^4s^3.
\end{eqnarray}
Since the initial Lagrangian $L$ implies the unique chain of four first class constraints, one expects that it has one local
symmetry of ${\stackrel{(3)}{\epsilon}}$-type. The symmetry can be found according to the defining
equations (\ref{25.2}), for the case
\begin{eqnarray}\label{E12}
\begin{array}{ccccc}
\epsilon^1 & +\dot\epsilon^2 & +2\epsilon^3\xi & {} & =0, \\
{} & 2\epsilon^2 & +\dot\epsilon^3 & +4\epsilon^4\xi & =0, \\
{} & {} & \epsilon^3 &  +\dot\epsilon^4 & =0.
\end{array}
\end{eqnarray}
It allows one to find $\epsilon^1, \epsilon^2, \epsilon^3$ in terms of $\epsilon^4\equiv\epsilon$:
$\epsilon^1=-\frac12{\stackrel{(3)}{\epsilon}}+4\dot\epsilon\xi+2\epsilon\dot\xi$,
$\epsilon^2=\frac12\ddot\epsilon-2\epsilon\xi$, $\epsilon^3=-\dot\epsilon$.
Using Eq. (\ref{25.1}), local symmetry of the Lagrangian (\ref{E1}) is given by
\begin{eqnarray}\label{E13}
\delta x^\mu=-\dot\epsilon x^\mu+2\epsilon\dot x^\mu, \qquad
\delta\xi=-\frac12{\stackrel{(3)}{\epsilon}}+4\dot\epsilon\xi+2\epsilon\dot\xi.
\end{eqnarray}
\par
\noindent
{\bf 2) Model with first and second class constraints.} Consider a theory with configuration space variables $x^\mu, ~ e, ~ g$
(where $\mu=0, 1, 2, 3, ~ \eta_{\mu \nu}=(-, +, +, +)$), and with action being
\begin{eqnarray}\label{E14}
S=\int d\tau\left(\frac{1}{2e}(\dot x^\mu-gx^\mu)^2+
\frac{g^2}{2e}\right), \qquad a=const.
\end{eqnarray}
One obtains the complete Hamiltonian
\begin{eqnarray}\label{E15}
H=\frac12 ep^2+g(xp)-
\frac{g^2}{2e}+v_{e}p_{e}+v_{g}p_{g},
\end{eqnarray}
as well as the constraints
\begin{eqnarray}\label{E16}
\Phi_1\equiv p_{e}=0, \qquad  T_1\equiv
-\frac12(p^2+\frac{g^2}{e^2})=0;
\end{eqnarray}
\begin{eqnarray}\label{E17}
\Phi_2\equiv p_g=0, \qquad \qquad T_2\equiv\frac{g}{e}-(xp)=0.
\end{eqnarray}
They can be reorganized with the aim to separate the first class constraints
\begin{eqnarray}\label{E18}
\tilde\Phi_1\equiv p_{e}+\frac{g}{e}p_{g}=0, \quad \tilde T_1\equiv
-\frac12(p^2-\frac{g^2}{e^2})-\frac{g}{e}(xp)+\frac{g^2}{e}p_{g}=0;
\end{eqnarray}
\begin{eqnarray}\label{E19}
p_g=0, \qquad \qquad \frac{g}{e}-(xp)=0.
\end{eqnarray}
The first (second) line represents the first (second) class subsets.

For the case, solution of the basic equation (\ref{9}) is given by
\begin{eqnarray}\label{E20}
\tilde p^\mu=\frac{1}{e-s^2}(\dot x^\mu-(g-s^2)x^\mu).
\end{eqnarray}
Using the equations (\ref{E16}), (\ref{E17}), (\ref{E20}) one obtains the extended Lagrangian (\ref{29.10})
\begin{eqnarray}\label{E21}
\tilde L=\frac{1}{2(e-s^1)}(\dot x^\mu-(g-s^2)x^\mu)^2+\frac{g^2}{2e}(1+\frac{s^1}{e})-\frac{g}{e}s^2.
\end{eqnarray}
Its local symmetries are obtained according to Eqs. (\ref{30}), (\ref{30.1}) using the expression (\ref{E18}) for the first class constraints
\begin{eqnarray}\label{E22}
\delta_1x^\mu=-\epsilon^1\left(\omega^\mu+\frac{g}{e}x^\mu\right), \qquad
\delta_1e=0, \qquad \delta_1g=\epsilon^1\frac{g^2}{e}, \cr
\delta_1s^1=\dot\epsilon^1-2\epsilon^1(\frac{gs^1}{e}-s^2), \qquad
\delta_1s^2=(\epsilon^1\frac{g}{e})^{\dot{}}+\epsilon^1\frac{g^2}{e};
\end{eqnarray}
\begin{eqnarray}\label{E23}
\delta_2x^\mu=0, \qquad \delta_2e=\epsilon^2, \qquad \delta_2g=\epsilon^2\frac{g}{e}, \cr
\delta_2s^1=\epsilon^2,  \qquad \delta_2s^2=\epsilon^2\frac{g}{e}.
\end{eqnarray}
Here $\omega^\mu$ is the r.h.s. of the equation (\ref{E20}).
By tedious computations one verifies that the variation $\delta_1\tilde L$ is the total derivative $\delta_1\tilde L$$=$$-\frac12(\epsilon^1(\omega^\mu)^2$$+\epsilon^1(\frac{g}{e})^2)^{\dot{}}$.

In the presence of second class constraints, local symmetries of $L$ can not be generally restored according to the trick (\ref{25.1}), (\ref{25.2}). The reason is that a number of equations of the system (\ref{25.2}) can be equal or more than the number of parameters $\epsilon^a$. In particular, for the present example one obtains just two equations for two parameters $\dot\epsilon^1$$+$$\epsilon^2$$=$$0$, $(\epsilon^1\frac{g}{e})^{\dot{}}$$+
$$\epsilon^1\frac{g^2}{e}$$+$$\epsilon^2\frac{g}{e}$$=$$0$.

{\bf 3) Maxwell action.} Consider the Maxwell action of electromagnetic field
%(on Minkowski space with the signature being $(-, +, +, +)$)
\begin{eqnarray}\label{E24}
S=-\frac14\int d^4xF_{\mu\nu}F^{\mu\nu}=\int d^4x\left[\frac12(\partial_0A_i-\partial_iA_0)^2-
\frac{1}{4}(F_{ij})^2\right].
\end{eqnarray}
For the case, the functions $v^i$ from Eq. (\ref{1}) are given by $p_i$ $+$ $\partial_iA_0$. The action implies primary and secondary constraints
\begin{eqnarray}\label{E25}
p_0=0, \qquad \partial_ip_i=0.
\end{eqnarray}
Then the basic equation (\ref{9}) acquires the form $\partial_0A_i$ $-$ $\omega_i$ $-$ $\partial_iA_0$ $+$ $\partial_is$ $=$ $0$, and the extended Lagrangian action (\ref{29.10})
is\footnote{In transition from mechanics to a field theory, derivatives are replaced by the variational derivatives. In particular, the last term in Eq. (\ref{9}) reads
$\frac{\delta}{\delta\omega_i(x)}$ $\int d^3ys^a(x)T_a(q^A(y), \omega_i(y)$.}
\begin{eqnarray}\label{E26}
\tilde S=\int d^4x\left[\frac12(\partial_0A_i-\partial_iA_0+\partial_is)^2-
\frac{1}{4}(F_{ij})^2\right].
\end{eqnarray}
Its local symmetries can be immediately written according to Eqs. (\ref{30.-1}), the nonvanishing variations are
\begin{eqnarray}\label{E27}
\delta_\beta A_0=\beta, \qquad \delta_\beta s=\beta, \cr
\delta_\alpha A_i=-\partial_i\alpha, \qquad \delta_\alpha s=\partial_0\alpha.
\end{eqnarray}
Symmetry of the initial action appears as the following combination
\begin{eqnarray}\label{E28}
(\delta_\beta+\delta_\alpha)A_i=-\partial_i\alpha, \cr
(\delta_\beta+\delta_\alpha)A_0=\beta,
\end{eqnarray}
where the parameters obey to the equation $\partial_0\alpha$ $+$ $\beta=0$. The substitution $\beta$$=$$-\partial_0\alpha$ into Eq. (\ref{E28}) gives the standard form of $U(1)$ gauge symmetry
\begin{eqnarray}\label{E29}
A'_\mu=A_\mu+\partial_\mu\alpha.
\end{eqnarray}

%with the Dirac procedure stopped on the third stage

\section{Conclusion}
In this work we have proposed an improvement of the extended Hamiltonian formalism for an arbitrary constrained system. Singular theory of a general form (with first and second class constraints of an arbitrary stage) can be reformulated as a theory that does not generate any constraints beyond the third stage. It is described by the extended Lagrangian constructed in terms of the original one according to Eq. (\ref{13}). All the higher-stage constraints of $L$ turn out to be the
second-stage constraints of $\tilde L$. The formalism implies an extension of the original configuration space $q^A$ by the auxiliary variables $s^a$. Number of them is equal to the number of all the higher stage constraints $T_a$ of original formulation. Those of the extra variables $s^a$ that correspond to the first class constraints, have been identified with the gauge fields supplying local symmetries of $\tilde L$. Hence in the passage from $L$ to $\tilde L$, local symmetries of $L$ with higher order derivatives of the local parameters decompose into a sum of the gauge type symmetries.

As an application of the extended Lagrangian formalism, we have presented a relatively simple way for obtaining the local symmetries of a singular Lagrangian theory.
By construction, the extended
Lagrangian implies only $\dot\epsilon$-type symmetries, that can be immediately written according to Eqs. (\ref{30}), (\ref{30.1}). The latter give the symmetries in terms of the first class constraints $\tilde G_{I_1}$ of the initial formulation and the coefficient functions of the constraint algebra (\ref{8}). Generators of transformations for all the original variables $q^A$ turn out to be the Lagrangian counterparts of canonical transformations generated by $\tilde G_{I_1}$. This result can be considered as a proof of the Dirac conjecture [1].
In contrast to a situation with symmetries of $L$ [14-16], the transformations (\ref{30}) do not involve the second class constraints.

The extended formulation can be appropriate tool for development of a general formalism for conversion
of second class constraints into the first class ones according to the ideas of the work [18]. To
apply the method proposed in [18], it is desirable to have the formulation with some configuration
space variables entering into the Lagrangian without derivatives. It is just
what happen in the extended formulation.

\section{Acknowledgments}
Author would like to thank the Brazilian foundations CNPq (Conselho Nacional de Desenvolvimento
Científico  e Tecnológico - Brasil) and FAPERJ for financial support.

\end{document}